\documentclass{IEEEtran}

\usepackage{booktabs} 
\usepackage{graphicx}
\usepackage{textcomp}
\usepackage{datetime}

\author{Jeffrey M. Shainline\\National Institute of Standards and Technology, 325 Broadway, Boulder, CO, 80305}

\begin{document}
\title{The largest cognitive systems will be optoelectronic}
\maketitle
\begin{abstract}
Electrons and photons offer complimentary strengths for information processing. Photons are excellent for communication, while electrons are superior for computation and memory. Cognition requires distributed computation to be communicated across the system for information integration. We present reasoning from neuroscience, network theory, and device physics supporting the conjecture that large-scale cognitive systems will benefit from electronic devices performing synaptic, dendritic, and neuronal information processing operating in conjunction with photonic communication. On the chip scale, integrated dielectric waveguides enable fan-out to thousands of connections. On the system scale, fiber and free-space optics can be employed. The largest cognitive systems will be limited by the distance light can travel during the period of a network oscillation. We calculate that optoelectronic networks the area of a large data center ($10^5$\,m$^2$) will be capable of system-wide information integration at $1$\,MHz. At frequencies of cortex-wide integration in the human brain ($4$\,Hz, theta band), optoelectronic systems could integrate information across the surface of the earth.
\end{abstract}

\section{\label{sec:intro}Introduction: cognitive systems}
Intelligent systems are characterized by the ability to incorporate a wide variety of stimuli into a coherent concept of the world \cite{enfr2001,vala2001,sase2001,bu2006,de2014,fr2015}. These stimuli generally encompass many content categories, and dynamically change across many temporal scales. Neural systems are excellent for cognition because they combine differentiated, local processing of neuronal assemblies \cite{budr2004,buwa2012} with information integration \cite{toed1998,tosp2003,to2004,seiz2006,bato2008,bato2009,base2011} across space \cite{busp2009,sp2010} and time \cite{enfr2001,vala2001,sase2001,bu2006,budr2004}. Small clusters of neurons code for certain features present in stimuli, and the neurons in these assemblies must communicate locally amongst themselves to form a consensus interpretation of a certain sensory input \cite{de2014}. This information must be communicated up the information-processing hierarchy and combined with input from other neuronal assemblies to form a broad, coherent, and multi-faceted conception of the available information \cite{bu2006,de2014,fr2015}. Information is integrated locally through transient synchronized oscillations at high frequencies, and information is integrated over larger regions of space through transient synchronization of larger numbers of neurons at lower frequencies \cite{stsa2000}. Cognition therefore depends on local computations amongst neuronal ensembles as well as communication locally and globally at various frequencies. Larger neural systems will be capable of more information processing, provided all neurons represent unique aspects of feature space, and information from all neurons can be integrated in a coherent cognitive state. In this work, we are concerned with the large-scale limits of cognitive systems.

It has been argued that neurons are uniquely suited to performing the differentiated processing and information integration across space and time required for cognition (see discussion in Ref.\,\cite{sh2018a}). Biological neurons perform the necessary information processing and memory operations using elegant electrochemical devices which grow in a bottom-up manner. The molecular-scale devices lead to very dense and compact circuits and systems. 

Many of the required neuronal functions can be performed with manufacturable electronic devices \cite{scpo2017}, such as transistors \cite{payu2017}, memristors \cite{huli2014,Cai2014}, magnetic tunnel junctions \cite{tori2018}, magnetic Josephson junctions (JJs) \cite{scdo2018}, and various superconducting devices \cite{shbu2017,sh2018b,sh2018c,sh2018d}. Such devices perform the necessary synaptic and neuronal processing which leads to network computation. By contrast, light is not naturally suited to perform neuronal computations. For example, neurons must sum the inputs from many connections. To perform this operation in the optical domain, a neuron would need to store photons in a cavity for long periods of time. Yet compact, high-$Q$ optical cavities with storage times longer than $1$\,ns are difficult to achieve, particularly when fabricated in an integrated process.

Biological neurons are excellent for computation, but communication via ionic conduction along axons is slow. This slow communication limits the total size of the neural system which can participate in synchronized oscillations at a given frequency, and therefore limits the total number of neurons that can be incorporated into a cognitive system \cite{bu2006}. Communication also poses challenges for semiconductor electronics. Fan-out in CMOS circuits as well as JJ circuits is limited to order 10. Light is not equipped for neuronal computations, but it is ideal for neuronal communication. Electrons and ions have charge and mass. They interact strongly and can be made to sit still. These traits lead to the potential for computation and memory. Photons are uncharged and massless. They do not interact, and they travel at the fastest velocity in the universe. On the local scale, the lack of interaction means photons in waveguides do not experience charge-based parasitics (resistance, capacitance, and inductance). A single optical pulse can fan out to as many ports as there are photons in the pulse without incurring an $RC$ penalty due to additional wiring. On the global scale, communication at the speed of light enables the largest area of neurons possible to be incorporated in a transient synchronized oscillation at a given frequency. By combining the strengths of electrons and photons, the information from many neurons with complex processing capabilities can be integrated across large regions of space to achieve the largest cognitive systems possible, given the light-speed limit of communication set by special relativity.

Based on graph theory metrics, we argue that even modestly sized networks of a few hundred thousand neurons require neurons to make a thousand synaptic connections on average. Already at the scale of networks of a few hundred thousand neurons, light provides significant advantages in communication. For large-scale systems comparable to the human brain, the advantages are more important. By considering the fundamental limit imposed by the light cone in conjunction with an estimate for artificial synapse area, we estimate the total number of synapses which can participate in a synchronized transient neuronal ensemble. We find that systems with ten billion times the number of synapses as a human brain may be possible. At these enormous scales, system power consumption is likely to be dominated by the production of light, so signaling with the fewest number of photons possible will reduce power density and total system power consumption. We argue that superconducting optoelectronic circuits are advantageous for enabling massive scaling with tractable power consumption. We close with speculation regarding the limits of cognitive systems using light for communication.

\section{\label{sec:differentiatedProcessing}Differentiated processing: neuronal assemblies}
\begin{figure}[h]
	\centerline{\includegraphics[width=8.6cm]{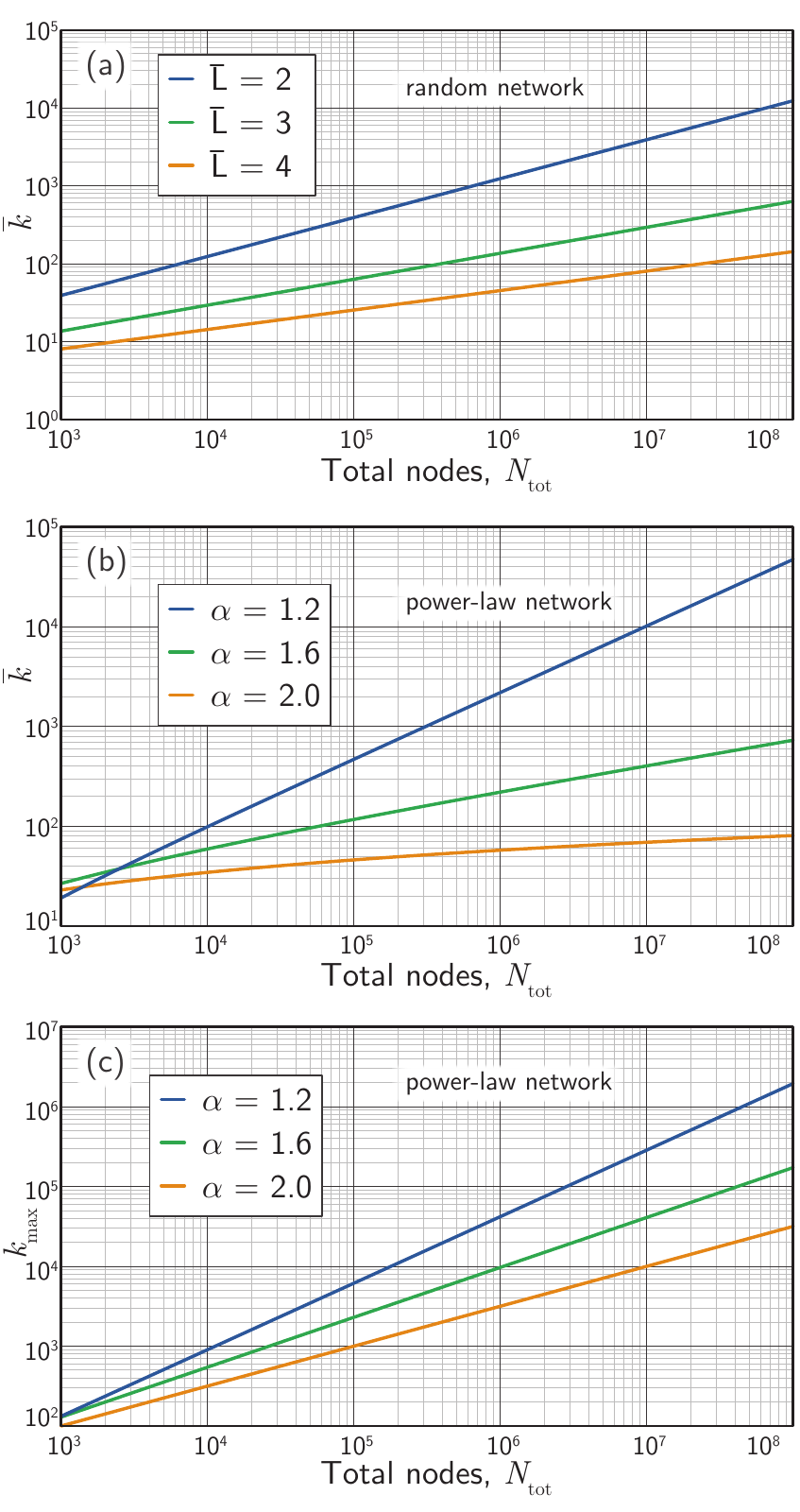}}
	\caption{\label{fig:networks_generalDegreeAnalysis}General degree analysis. (a) Average degree of a node in the network as a function of total number of nodes in the network for a random network for three values of average path length. (b) Average degree of a node in a network as a function of the total number of nodes in the network for a power-law degree distribution. (c) Maximum degree of a single node in a network as a function of the total number of nodes in the network for a power-law degree distribution.}
\end{figure}
Neural systems represent information in the firing rate and relative timing of neurons. A given neuron can represent a limited subset of all possible stimuli, and clusters of neurons form transient assemblies that temporarily synchronize to form a consensus interpretation of a given input. The information from many assemblies must be communicated up the cognitive hierarchy in a bottom-up manner, and feedback must also be provided from higher cognitive centers to local neuronal ensembles in a top-down manner to provide information transfer across the system. For communication to be efficient across the network, each node must be able to send and receive information to and from many other nodes. In the language of network theory \cite{eskn2015}, communication will be efficient if the average path length \cite{frfr2004} of the network is short. The average path length is determined by calculating the smallest number of steps from each node of the graph to every other node, and averaging this quantity over all pairs of nodes. In order to represent large quantities of information, we would like a network with as many neurons as possible, and for efficient integration of the information throughout the network, we need short path length. We therefore wish to investigate path length in networks which are useful for cognition.

To identify networks useful for cognition, we need to analyze connectivity. An important network metric related to connectivity is the degree distribution. The degree of a node refers to the number of connections (synapses) it forms with other nodes in the network. The degree distribution refers to the statistical distribution of the degrees of all the nodes in the network. We consider two degree distributions that are pertinent to information processing in neural systems. A random network \cite{eskn2015} is characterized by a Gaussian degree distribution, and a scale-free network \cite{baal1999,egch2005} is characterized by a power-law degree distribution of the form $p(k) \propto k^{-\alpha}$, where $p(k)dk$ is the probability of finding a node with degree between $k$ and $k+\delta k$, and $\alpha$ is a constant, usually between 1 and 3. Random networks, formed by assigning connections at random between pairs of nodes, achieve short path lengths between any two nodes in the network, and are effective at forming associative memories, as observed in hippocampus \cite{bu2006}. Scale-free networks arise due to the growth conditions in many natural contexts \cite{baal1999}, and appear to fit the connectivity of large-scale brain networks \cite{egch2005}. Recent work finds evidence that degree distributions of neurons in human cortex are fit well by stretched exponential functions \cite{gaod2016}. Here we consider power-law and Gaussian degree distributions, keeping in mind that stretched exponentials or other degree distributions may better model cortex. The general conclusions drawn by this analysis are not likely to be affected by the exact form of the underlying degree distribution.

To investigate the potential for efficient information integration in neuronal assemblies of various sizes, we investigate the relation between the number of neurons in the network and the number of connections made by the neurons for Gaussian and power-law degree distributions. For the Gaussian degree distribution of a random network, the average degree of a node in the network, $\bar{k}$, is related to the total number of nodes in the network, $N_{\mathrm{tot}}$, and the average path length, $\bar{L}$, by the relation $\bar{k} = \mathrm{exp}\{[\mathrm{ln}(N_{\mathrm{tot}})-\gamma]/\bar{L}+1/2\}$ \cite{frfr2004}, where $\gamma \approx 0.5772$ is Euler's constant. This relationship between the total number of nodes in the network and their degree is shown in Fig.\,\ref{fig:networks_generalDegreeAnalysis}(a) for three values of $\bar{L}$. From this plot we find that for modest networks of 100,000 nodes, the average degree must be 400 to ensure an average path length of two, and for a network of one million nodes, the average degree must be over one thousand to keep the path length to two. In hippocampus, the average path length is believed to be less than two \cite{bu2006}, and neurons are observed to have over ten thousand connections. This extraordinary connectivity appears to be necessary to enable memory recall (convergence to an attractor \cite{st2015}) within a single cycle of gamma oscillation \cite{alid2007}. More generally, large cognitive systems are likely to make use of modules with large numbers of neurons with random connectivity to form associative memories with large storage (like hippocampus), and these modules are likely to need short path length for efficient information access within a small number of network oscillations.

Figure\,\ref{fig:networks_generalDegreeAnalysis}(b) presents a similar analysis for the case of the power-law-distributed network. Here we plot the average degree versus the total number of nodes in the network for three values of the exponent, $\alpha$. When viewing Figs.\,\ref{fig:networks_generalDegreeAnalysis}(b) and (c), one should keep in mind that smaller $\alpha$ corresponds to shorter path length because more high-degree nodes are included in the network. The node of the network with maximum degree, $k_{\mathrm{max}}$, is shown in Fig.\,\ref{fig:networks_generalDegreeAnalysis}(c). Again we see that modest network sizes require very high degree nodes. The message of Fig.\,\ref{fig:networks_generalDegreeAnalysis} is that information integration in cognitive neural systems requires massive connectivity. This massive connectivity is necessary to enable rapid information access and integration across neuronal ensembles embedded within cognitive modules. Neurons in the brain making tens of thousands of connections across multiple regions of the thalamocortical complex with long-range axonal projections are central to achieving the information integration necessary for cognition \cite{de2014}. As we consider the limits of cognition based on physics, we must consider how to best achieve neurons with massive connectivity without sacrificing the potential for information integration in the temporal domain.

\section{\label{sec:electricalAndOptical}Hardware: electrical and optical systems}
The connectivity requirements of neurons have been widely appreciated for quite some time, and the majority of neuromorphic engineers working with CMOS systems decide not to design neurons with such a large number of direct connections. Direct wiring would introduce large capacitance and resistance, which would slow response times and require too much current and power. Instead, virtual connections are achieved by assigning addresses to neurons, and shared communication lines route synaptic events between neurons based on addresses in a synaptic routing table. When multiple events request access to the shared communication lines, the events are queued, and latency related to the number of neurons on the bus and their firing frequencies is introduced. Communication over shared lines is referred to as time multiplexing, and it is a common and powerful technique with CMOS hardware across multiple computing domains. The speed of semiconductor devices mitigates limitations in connectivity \cite{bo2000}. 

Time multiplexing signals with address-event representation has achieved some of the most impressive neuromorphic systems to date \cite{payu2017,bega2014,mear2014,frsc2017,dasr2018}. The functionality of CMOS systems enables reconfigurable networks with various forms of synaptic plasticity and dendritic processing, illustrating the fitness of electronics for performing neuronal computations. Yet the challenges inherent in communicating with electrons are evident in neuromorphic CMOS. The necessity of shared communication lines and routing nodes introduces a constraint on the number of neurons participating in a synchronized ensemble and the oscillation frequency of that ensemble. In silicon microelectronics, time-multiplexed communication lines are necessary between nodes above a certain degree due to the immutable physical properties of electrons. The high speed of CMOS devices enables impressive neuromorphic systems despite communication bottlenecks. Yet we should not let the excellence of CMOS deter us from considering physical systems even more equipped for cognition.

Neurons in a physical system without charge-based parasitics can achieve direct fan-out to thousands of synaptic connections, avoiding the need for time-multiplexing necessitated by shared communication lines. By using photons rather than electrons for communication across multi-planar routing structures \cite{chbu2017,sami2017}, 10-to-100 routing manifolds have been demonstrated \cite{chbu2018}, and direct fan-out to thousands of connections appears straightforward. By utilizing optical fibers and free-space links in addition to on-chip routing networks, signaling across large-scale, multi-modular cognitive systems may be achieved. 

Communication appears promising in emerging integrated-photonic systems \cite{suwa2015,atmo2018}, but inevitably, new challenges are introduced. First, light sources are inefficient, and it is not worth the energy cost to produce light if only a small number of connections will be addressed. Second, using light for communication requires hardware with light sources and detectors integrated with the electronic neuronal components. The challenge of constructing this hardware is the most significant near-term hurdle to overcome on the route to large-scale optoelectronic cognitive systems. Third, the relatively large wavelength of light makes optical components generally larger than electronic components (and far larger than biological neuronal components). Models of optical neurons indicate they may be roughly the same size as CMOS neurons \cite{payu2017,sh2018e}, and as we argue below, the factor determining the scaling potential of a cognitive system scales as the velocity of communication divided by the size of a neuron or synapse. It appears light speed communication more than compensates for the increase in device size of optical neurons relative to biological neurons. The large wavelength of light is an inconvenience, but it does not appear to negate the strengths of light for communication in large-scale cognitive systems. As optoelectronic hardware matures, the progress made to overcome these challenges will determine the scale at which light for communication becomes advantageous.

In addition to spatial and graph metrics, there are many considerations at the level of devices, circuits, and architecture that are necessary to assess utility. Neuronal devices and circuits must achieve excitatory and inhibitory connections with dendritic processing to maximize utilization of the frequency domain. For very large neural systems, unsupervised learning is required for scalability, so a variety of synaptic plasticity mechanisms are required to enable memory retention and learning in the presence of dynamically varying stimuli. Networks and systems must achieve small-world graph metrics and information processing across interconnected modules to enable efficient exchange between sub-processors. Optoelectronic neurons and networks designed with these criteria in mind have been presented in Refs.\,\cite{sh2018a,sh2018b,sh2018c,sh2018d,sh2018e}. While significant further investigation of the requirements on neurons and network architectures for cognition is required, the present work proceeds assuming that electronic circuits (either semiconducting or superconducting) can achieve a wide variety of necessary device functions, and optical networks can achieve arbitrary connectivity graphs. Future knowledge regarding the requirements for cognition are likely to lead to modifications of circuit and system designs, but the conjecture that electronic circuits will remain capable of performing neuronal and synaptic functions while photonic networks will remain capable of performing communication is likely to hold. 

\section{\label{sec:informationIntegration}Information integration: the neuronal pool}
We have argued that electronics are excellent for performing the synaptic, dendritic, and neuronal operations required by cognitive systems, and also that light is advantageous for enabling direct fan-out to thousands of connections, thereby circumventing the trade-offs introduced by shared electronic communication lines. The advantage in communication at the chip scale results from the absence of charge-based parasitics, but for communication across large scales (across a wafer, between wafers in a module, or between multi-wafer modules in a massive cognitive system), the speed of light plays a central role in facilitating the highest performance neuronal operation.

\begin{figure}[h]
	\centerline{\includegraphics[width=8.6cm]{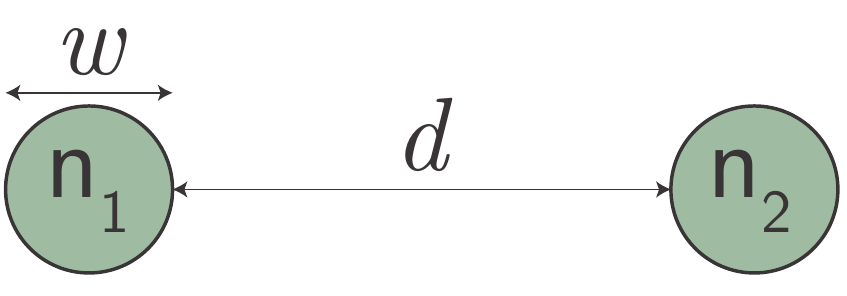}}
	\caption{\label{fig:neuronalPool}Schematic of two neurons of width $w$ separated by distance $d$ considered in the calculation of the number of neurons in the neuronal pool.}
\end{figure}
As discussed in Sec.\,\ref{sec:intro}, neural systems integrate information through transient synchronized oscillations across a multitude of spatial and temporal scales. To anticipate very large cognitive systems, we consider the number of neurons and synapses that can be integrated in a transient synchronized oscillation at a given frequency. We refer to an ensemble of neurons capable of synchronizing as the neuronal pool. Consider two neurons, each of width $w$, separated by a distance $d$, as shown in Fig.\,\ref{fig:neuronalPool}. If $\mathsf{n_1}$ produces a spike at time $t_0$, and $\mathsf{n_2}$ produces a spike at $t_0+\Delta t$, we ask whether the spike produced by $\mathsf{n_1}$ may have contributed to the spike produced by $\mathsf{n_2}$, and therefore whether $\mathsf{n_1}$ is capable of inducing $\mathsf{n_2}$ to synchronize within a single period of the oscillation cycle. If the neurons communicate with signals propagating at velocity $v$, the pulse from $\mathsf{n_1}$ will have traveled a distance $x = v \Delta t$ by the time $\mathsf{n_2}$ spikes. For the case of synchronized oscillations at frequency $f$, the oscillation period $T = 1/f$ sets the time scale. We find that if 
\begin{equation}
\label{eq:diameterOfPool}
d \le \frac{v}{f},
\end{equation} 
$\mathsf{n_1}$ may induce $\mathsf{n_2}$ to spike within a single period of oscillation. We take the value of $d$ which saturates the inequality of Eq. \ref{eq:diameterOfPool} to define the maximum diameter of the neuronal pool. 

One may argue that to establish synchronization, $\mathsf{n_1}$ must induce $\mathsf{n_2}$ to spike, and this spike from $\mathsf{n_2}$ must also propagate to $\mathsf{n_1}$, causing $\mathsf{n_1}$ to spike again during the next period of the oscillation cycle. This model would decrease $d$ by a factor of two. One may also argue that $\mathsf{n_1}$ could produce a spike train at frequency $f$, and this spike train could induce $\mathsf{n_2}$ to produce a spike train of the same frequency at a later time, independent of their spatial separation. In the context of cognitive processing, the delayed pulse train from $\mathsf{n_2}$ is not conducive to efficient information integration if it is delayed beyond a single oscillation period. Much like networks with short path length are necessary for efficient information integration across space, signaling with delay shorter than a single period of oscillation is necessary for information integration across time. For this reason, we consider the diameter of the neuronal pool to be set by the distance signals can propagate within a single period of oscillation.

It is not the diameter of the neuronal pool that is of primary interest, but rather the number of neurons (or synapses) that can integrate information within the pool. To estimate this quantity, we must account for the size of a neuron, labeled $w$ in Fig.\,\ref{fig:neuronalPool}. The number of neurons in the pool, $N_{\mathrm{pool}}$, is given by the size of the pool divided by the size of a neuron. In $n$ dimensions we have 
\begin{equation}
\label{eq:numberInPool}
N_{\mathrm{pool}} = \left(\frac{v}{wf}\right)^n.
\end{equation}

\begin{figure}[h]
	\centerline{\includegraphics[width=8.6cm]{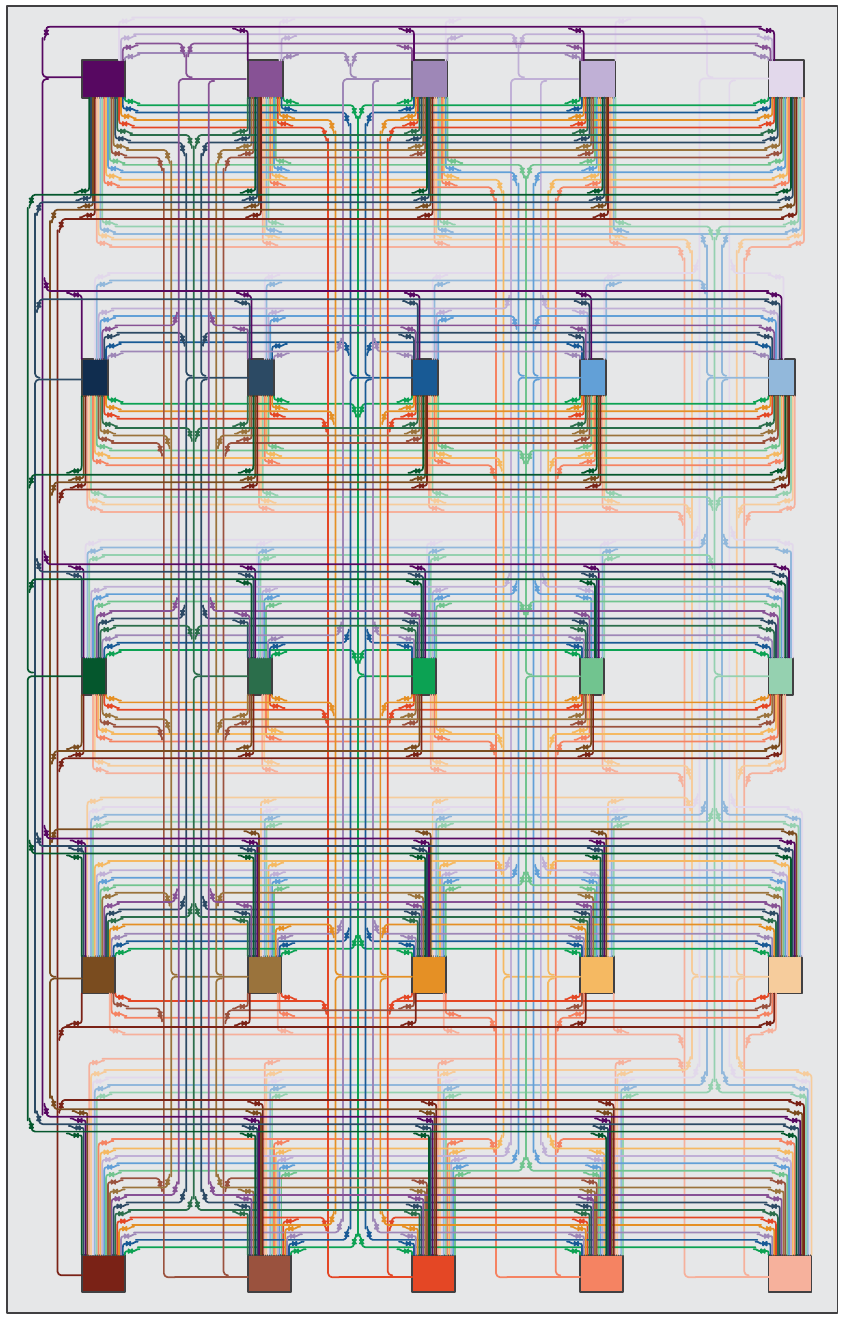}}
	\caption{\label{fig:networks_routingDiagram_full_allToAll}Row-column routing architecture in 5 $\times$ 5 sector. All-to-all connections.}
\end{figure}

\noindent Note that when we refer to the neuronal pool, we do not refer to the number of neurons which are synchronized at a given moment, which is a quantity that depends on the network graph and dynamical state. In this calculation, we refer to the total number of neurons that could potentially synchronize based on the reach of communication.

\begin{figure}[h]
	\centerline{\includegraphics[width=8.6cm]{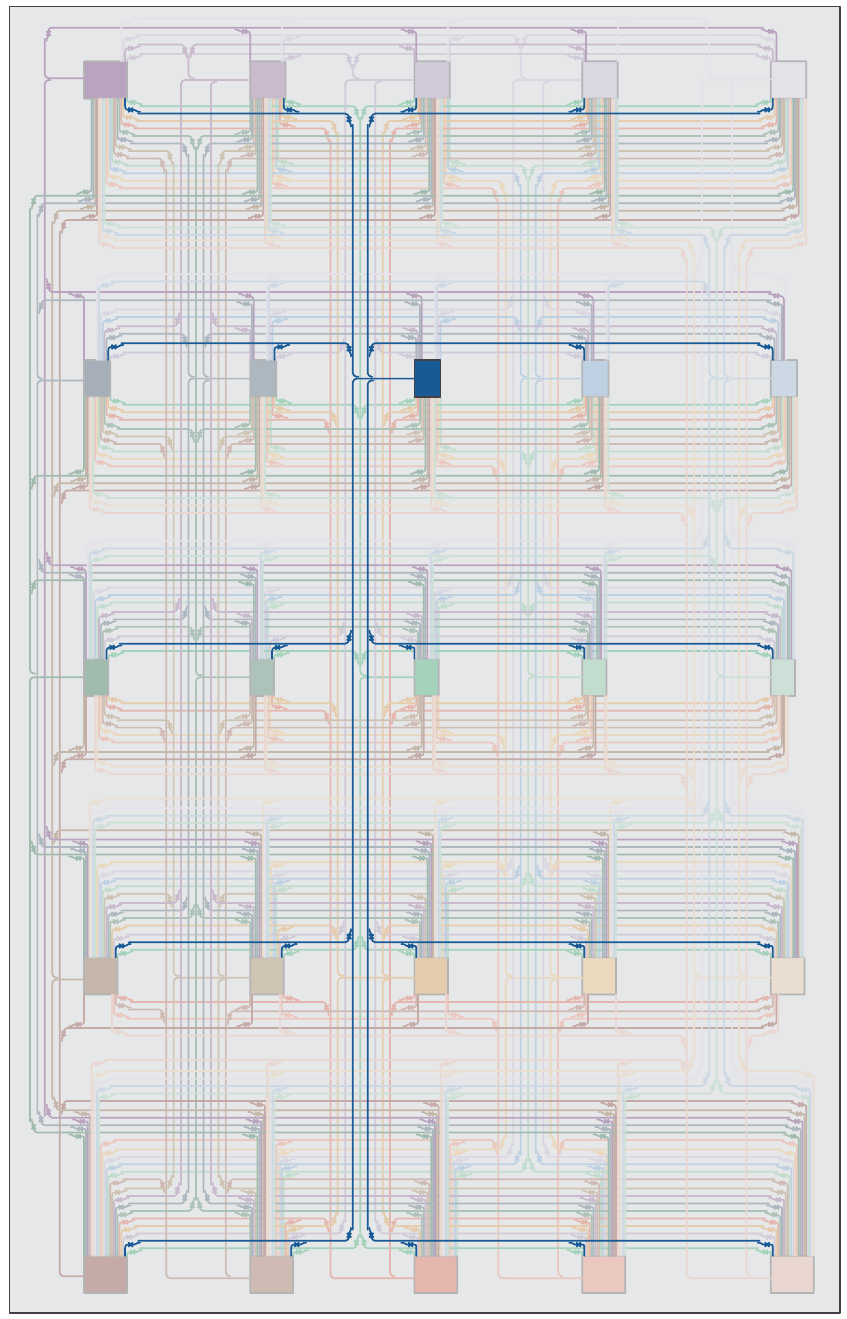}}
	\caption{\label{fig:networks_routingDiagram_full_oneToAll}Row-column routing architecture in 5 $\times$ 5 sector.  The out-directed connections from a single node are highlighted.}
\end{figure}
Let us check the validity of Eq.\,\ref{eq:numberInPool} with numbers from the human brain. Cortex in mammals is a two-dimensional sheet. When removed from the skull, it can be unfurled flat on a surface. Therefore, we model signals propagating within a plane, and take $n=2$ for this analysis, although it may be possible to construct artificial cognitive systems fully utilizing three spatial dimensions. The area of the human cerebral cortex is $0.095$\,m$^2$ \cite{scva2014}, and it contains $1.6\times10^{10}$ neurons \cite{azca2009,he2009}, giving $w = 2.4\times 10^{-6}$\,m. Signal velocity along axons in cortex is $2$\,m/s \cite{edve2004}, and theta oscillations ($4$\,Hz-$8$\,Hz) are believed to play a key role in integrating information across cortex. Using $v = 2$\,m/s, $w=2.4\times 10^{-6}$, and $f = 6$\,Hz in Eq. \ref{eq:numberInPool}, we find $N_{\mathrm{pool}} = 1.9\times 10^{10}$ neurons, very close to the reported value of $1.6\times 10^{10}$ neurons \cite{azca2009,he2009} in the human cerebral cortex.

\begin{figure}[h]
	\centerline{\includegraphics[width=8.6cm]{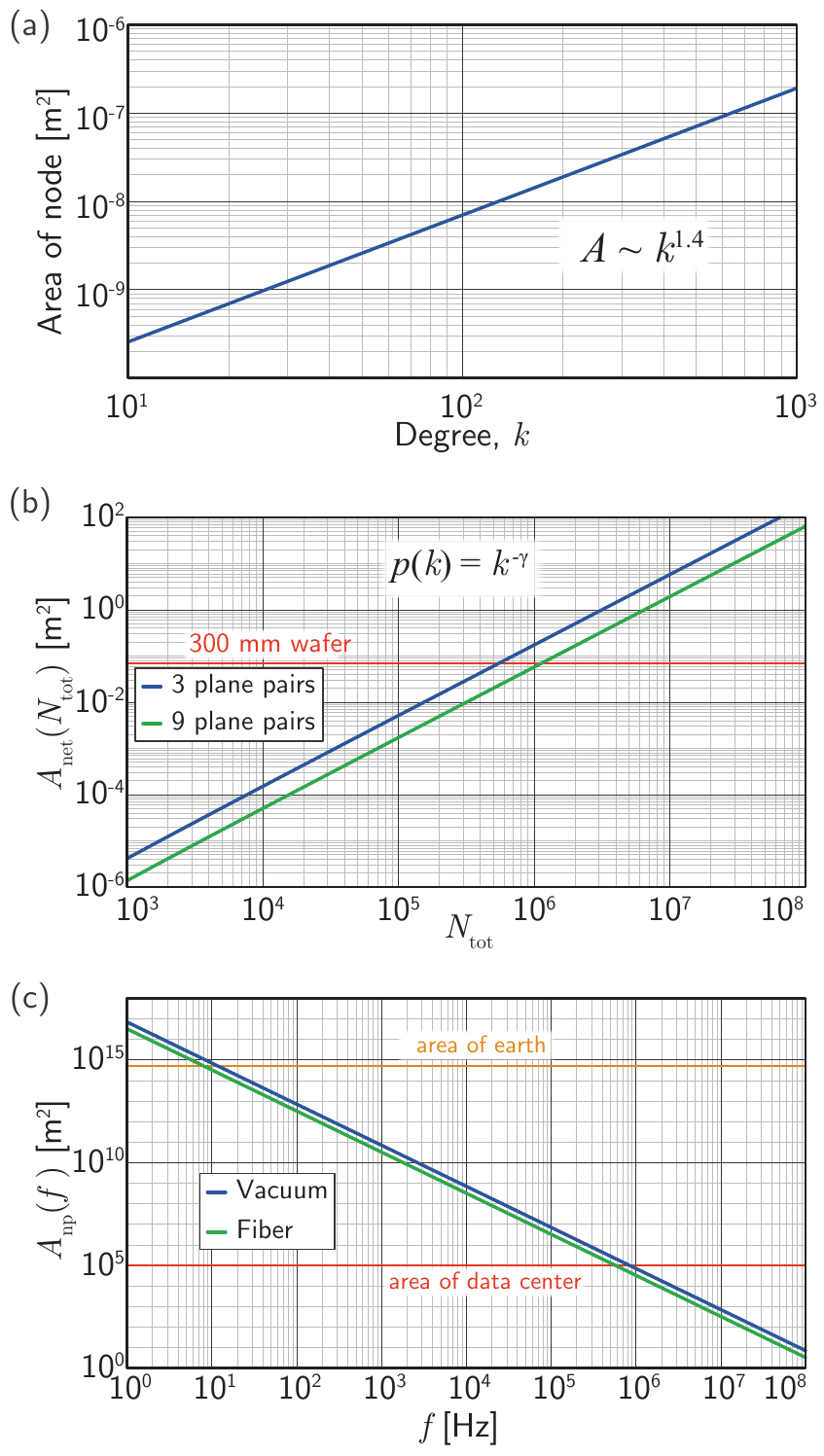}}
	\caption{\label{fig:networks_scaling}Spatial scaling of optoelectronic networks. (a) Area of a node as a function of its degree based on the routing architecture shown in Fig.\,\ref{fig:networks_routingDiagram_full_allToAll} and the device designs presented in Refs.\,\cite{sh2018b,sh2018c,sh2018d,sh2018e}. (b) Area of the network as a function of the total number of nodes assuming a power-law degree distribution. (c) Area of the neuronal pool as a function of the frequency of network oscillations, assuming light-speed communication.}
\end{figure}

To assess the performance of a neural system using light for communication, we must estimate the size of optoelectronic neurons. Such a spatial estimate requires an assessment of the size of the neurons themselves as well as the space occupied by routing waveguides, analogous to white matter in the brain. This analysis is conducted in Ref.\,\cite{sh2018e}. A schematic of the proposed routing architecture is shown in Fig.\,\ref{fig:networks_routingDiagram_full_allToAll}. This routing schematic leverages multiple planes of integrated photonic waveguides \cite{chbu2017,sami2017,chbu2018} to achieve all-to-all connectivity, and networks in practice can be constructed by pruning the all-to-all master routing scheme. The connections emanating from a single node are highlighted for clarity in Fig.\,\ref{fig:networks_routingDiagram_full_oneToAll}. Based on this routing and analysis of the area required for synaptic and neuronal circuits \cite{sh2018b,sh2018c,sh2018d,sh2018e}, we calculate the area of a neuron as a function of its degree, as shown in Fig.\,\ref{fig:networks_scaling}(a). Assuming a power-law degree distribution, we calculate the total area of the network, as shown in Fig.\,\ref{fig:networks_scaling}(b). We find a network with one million neurons and two hundred million synapses will fit on a 300\,mm wafer. Further details are included in Ref.\,\cite{sh2018e}.

To compare two hardware platforms, we consider spatial scaling in terms of the size of a synapse rather than a neuron. Neurons many have any number of synapses, and therefore may span a wide range of sizes, whereas the size of synapses will depend less on the degree of the neuron to which they are connected. In particular, we wish to compare neurons leveraging photonic communication to the human brain. Denoting the photonic and biological hardware platforms with superscripts, we find
\begin{equation}
\label{eq:neuronalPool_number}
\frac{N_{\mathrm{pool}}^{(\mathrm{p})}}{N_{\mathrm{pool}}^{(\mathrm{b})}} = \left(\frac{v^{(\mathrm{p})}w^{(\mathrm{b})}}{w^{(\mathrm{p})}v^{(\mathrm{b})}}\right)^2.
\end{equation}
To be concrete, we consider the size of superconducting optoelectronic neurons, as described in Ref.\,\cite{sh2018e}. A 300 mm wafer can support roughly $2.0\times 10^8$ superconducting optoelectronic synapses, giving an estimate of $w^{(\mathrm{p})} = 1.9\times10^{-5}$\,m. Here $w$ refers to the size of a synapse, not a neuron, as discussed above. If we assume each of the $1.6\times 10^{10}$ neurons in the human cerebral cortex has $10^4$ synapses \cite{brsc1998}, we find $w^{(\mathrm{b})} = 2.4\times 10^{-8}$\,m. A biological synapse is 1000 times smaller than a superconducting optoelectronic synapse in width, and a million times smaller in area. The speed of signals in cortex is roughly 2\,m/s \cite{edve2004}. Axons with larger diameter can propagate signals above 100\,m/s, but for the dense connectivity of cortex, such large fibers cannot be supported. The speed of light is $3\times 10^8$\,m/s \cite{speedOfLight}. Thus, comparing superconducting optoelectronic networks to biological networks, we find $N_p^{(\mathrm{p})}/N_p^{(\mathrm{b})} \approx 10^{10}$. The neuronal pool enabled by light-speed communication can contain ten billion times the number of synapses as the pool enabled by ionic signal propagation along biological axons. Signaling at the speed of light brings a tremendous advantage in this regard, an advantage made more significant in networks spanning a volume rather than an area. This simple scaling analysis does not take into account factors that may be significant for very large systems, such as volume required for liquid helium flow for cooling or the volume of white matter occupied by optical fibers carrying signals between large modules. Suppose these factors introduce a quadratic error, and the correct scaling is the square root of the $10^{10}$ estimate. We would still be considering a system with $10^5$ times the number of synapses as the human brain, and orders of magnitude faster network oscillations.

We can calculate the area of the neuronal pool as a function of the frequency of oscillations, as shown in Fig.\,\ref{fig:networks_scaling}(c). We find that oscillations at $1$\,MHz can integrate information across an area of $10^5$\,m, roughly the size of a large data center. For the optoelectronic hardware considered in Refs.\,\cite{sh2018a,sh2018b,sh2018c,sh2018d,sh2018e}, oscillation frequencies above tens of megahertz will be attainable, and therefore such systems could integrate information from across a system the size of a data center into a coherent cognitive state within 1\,\textmu s. If system-wide oscillations occurred at the same frequencies as the human brain (theta band, 4\,Hz - 8\,Hz), the network could be as large as the earth, integrating the information within local regions of $10^5$\,m$^2$ every microsecond, and integrating the information from thousands of these regions several times per second.

On the small scale of a chip, photonic communication enables dense local clusters of optoelectronic neurons by direct signal fan-out. On the large scale of multi-modular systems, photonic communication enables information integration across massive cognitive systems by communicating at the fastest velocity in the universe. The strengths of electronics for computation and photonics for communication indicate that the largest cognitive systems will be optoelectronic.

\section{\label{sec:power}Power: as few photons as possible}
Once optical communication is employed, communication velocity is not likely to be the factor which limits network size in the near term, as it appears to be for biological systems. When we are discussing networks as big as data centers, power consumption becomes the center of attention. Assuming communication requires creation and destruction of a photon, it is not possible to send optical signals with less energy than a single photon. We therefore conjecture that neurons communicating to their synaptic connections with order one photon will achieve the highest energy efficiency. 

One must consider which frequency of photons are best suited for this application. Because the energy of a photon and its wavelength are inversely proportional, optoelectronic circuits face a power/area trade-off. It may be possible to develop circuits based on microwave photons to achieve extreme energy efficiency. However, for integrated circuits, the long wavelength of microwave photons may make dense integration of highly connected chip- and wafer-scale networks difficult. Alternatively, ultraviolet photons could be employed to reduce the device footprint. In this case, materials are less conducive to integrated fabrication, and photon production requires an undesirable amount of energy. At present, it appears that near-infrared photons ($\lambda\approx$\,1.5\,\textmu m) strike a balance between size and energy efficiency. Additionally, based on the ubiquity of near-infrared light sources (including silicon-based \cite{shxu2007,buch2017}); the integrability \cite{shbu2017b} and performance \cite{mave2013} of near-infrared single-photon detectors; and the abundance of passive waveguiding media (including optical fibers), this wavelength range appears uniquely suited to achieve the highly scaled optoelectronic systems under consideration.

Superconducting single-photon detectors \cite{gook2001,nata2012,liyo2013,mave2013} will perform well receiving communication events between neurons. The use of superconductors in large-scale cognitive systems contributes to energy efficiency by enabling single-photon communication and also because the superconducting circuits draw near zero power when not responding to a detection event. Further, using single-photon detectors in conjunction with other superconducting electronic circuit elements, such as Josephson junctions \cite{ti1996,vatu1998,ka1999} and thin-film amplifiers \cite{mcbe2014,mcab2016,zhto2018}, combines the strengths of photons for communication with electronic devices for computation and memory. Superconducting optoelectronic circuits achieving the desired synaptic and neuronal functions have been designed in Refs.\,\cite{sh2018b,sh2018c,sh2018d}. Future designs achieving communication without photon annihilation, or neurons leveraging concepts of reversible computing may achieve further energy efficiency.

As we have argued, neurons with many long-range connections are important for information integration across a cognitive system. In the context of an optoelectronic hardware platform, consider a massive node with $10^6$ synaptic connections spanning multiple regions of the network. Suppose 10 photons per firing event must be sent to each synapse to overcome loss and noise. A pulse of $10^7$ photons consumes a picojoule at $\lambda = 1.5$\,\textmu m. Even with a poor photon production efficiency of $10^{-3}$, this massive, long-range neuron could synchronize vast neuronal ensembles across multiple processing areas at 1\,MHz with with 1\,mW of device power. This large node would address an order of magnitude more synapses than the largest long-range neurons in the human brain.

The use of superconductors contributes to energy efficiency by enabling communication with single photons, but superconductors require cryogenic operation, which is inefficient. Is there a net gain in energy efficiency? For large systems, the answer is likely yes. To operate at liquid helium temperature, 4.2\,K, approximately one kilowatt of cooling power is required for each watt of device power. Cryogenic operation of sources can gain two orders of magnitude in efficiency \cite{doro2017}. Similarly, low-temperature waveguide-integrated superconducting detectors efficiently receive single photons, while room-temperature waveguide-integrated, scalable semiconductor photodetectors may require one thousand photons or more. While cryogenic operation costs three orders of magnitude due to power required for cooling, the combined improvements in sources and detectors may gain five orders of magnitude in power consumed by devices. Noise will also be much higher at room temperature when faint photonic signals are employed. Additionally, the extraordinarily low energy per operation of Josephson electronics leads to synaptic and dendritic computations with extremely low power. Calculations of a network of one million neurons in operation with oscillations up to 20\,MHz indicate the network will consume 1\,W, with a power density of 10\,W/m$^2$ \cite{sh2018e}. Considering spatial scaling in conjunction with power scaling, we find this value of 10\,W/m$^2$ is roughly constant for networks across a wide range of spatial scales. The heat from these networks can be straightforwardly removed with liquid helium \cite{ek2006}, indicating cooling will not be a fundamental impediment when scaling to large cognitive systems.

While the arguments related to communication and energy efficiency lead us to conjecture that superconducting optoelectronic networks will be uniquely suited to achieving large-scale cognitive systems, we emphasize that room-temperature neural systems will also benefit from the strengths of optoelectronic integration. Operation by humanoid organisms in an earth-like environment necessitates that at least some portion of the system operate in an ambient environment. The advantages of using light for communication across cognitive systems lead us to anticipate hybrid cognitive systems with high-speed, high-efficiency cognition occurring in cryogenic modules, and information integration to room temperature with optical signals over fiber. The advantages of light for communication will be paramount for enabling high-bandwidth, low-latency signaling between modules in a massive cognitive system, spanning cryogenic and ambient environments. 

\section{\label{sec:outlook}Scaling: cognition across the solar system?}
How large can optoelectronic cognitive systems be? As we see in Fig.\,\ref{fig:networks_scaling}(c), networks as large as the earth can communicate with oscillations near 1\,Hz. Earth is the most valuable environment in the solar system to humans because of its magnetic field, atmosphere, and temperature. These attributes make the earth hospitable to humans, but are not necessary for optoelectronic cognitive systems. Extraterrestrial objects are naturally suited to support these systems. In particular, asteroids are abundant, sufficiently large, colder, and composed of the necessary silicates and superconductors to constitute the large-scale optoelectronic systems under consideration \cite{mufo2017,astra,bu1999,shcl2010,necl2014}. Information could be integrated across an asteroid with diameter of 60\,km \cite{hada2015} through transient synchronized oscillations at $1$\,kHz. 

It appears possible for an asteroid belt to form the nodes and light to form the edges of a solar-system-scale intelligent network. Would this turn the entire solar system into a single cognitive module? Probably not. Asteroids can be separated by billions of meters, so light-speed communication delays may be several seconds or longer. For cognitive systems oscillating from $1$\,kHz to $20$\,MHz, such delays would cause individual modules to operate as separate cognitive units. Light-speed communication between these modules would be analogous to spoken words between humans, and the asteroid belt would form a community of cognitive entities.

What types of computations might these systems conduct? Allow us to speculate. Superconducting technologies are capable of digital logic \cite{li2012,taoz2013,hehe2011} and quantum computing \cite{we2017}. Quantum state behavior is fundamentally probabilistic, and neural systems are optimized for statistical inference \cite{mabe2006,yash2007,bema2008}. Cognitive neural systems coupled to quantum systems may engage in computation that leverages quantum superposition and entanglement without necessitating high-fidelity gates, long qubit lifetimes, or error correction. Superconducting optoelectronic hardware may enable this type of computation, with data sent to and from the cryogenic environment over optical fibers, quite likely received in the ambient environment by optoelectronic CMOS, perhaps augmented by distributed memory and nanoscale dynamical elements \cite{tori2018}. Further, superconducting sensors \cite{alve2015,chsc2017}, including single-photon detectors, are excellent for detection of radiation of many types at many energies. Systems leveraging these sensors are presently used in exoplanet search \cite{raca2016,boga1992,kila2016}, cosmology \cite{diad2017}, and particle detectors \cite{le2017}. Cognitive systems working in conjunction with a variety of superconducting sensors have the potential to observe the universe across large scales of space and time.  A cognitive system receiving stimulus from many other solar systems and galaxies reaching to the edge of the observable universe and computing with classical, quantum, and neural information may have insights into the origin and evolution of nature beyond what we can currently conceive.

\section{\label{sec:summary}Summary: photons and electrons for cognition}
From neuroscience we know that a variety of neuronal processors and assemblies send information up the cognitive hierarchy, while neurons and systems of neurons at higher-levels of hierarchy send feedback across the network to close processing loops and inform local clusters of the consensus being established by the entire cognitive system. Large neurons with massive connectivity are crucial for network information integration, and for the fastest convergence to a network consensus, neurons across large regions must be able to drive each other to fire within the period of a network cycle to efficiently share information across space and time. Electronic circuits are excellent for local, synaptic, dendritic, and neuronal processing. Photonic circuits are excellent for dense, local connectivity, as well as long-range communication. These arguments lead to the conjecture that tremendous gains in cognition will be enabled by advances in optoelectronic hardware.

\vspace{0.5em}
This is a contribution of NIST, an agency of the US government, not subject to copyright.

\bibliographystyle{IEEEtran}
\bibliography{OptoelectronicCognition}

\end{document}